\documentclass[aps,prl,reprint,superscriptaddress,nofootinbib,showpacs,preprintnumbers]{revtex4-1}
\pdfoutput=1

\usepackage{graphicx} 
\usepackage{subfigure}
\usepackage{longtable}
\usepackage{multirow}
\usepackage{textcomp}
\usepackage{units}
\usepackage{rotating}
\usepackage{amssymb}
\usepackage{amsmath}
\usepackage{array} 
\usepackage[hyperindex,breaklinks]{hyperref}
\DeclareGraphicsExtensions{.pdf}

\usepackage{atlasphysics}
\usepackage{preprintcover}

\PreprintCoverPaperTitle{
Determination of the strange quark density of the proton from 
ATLAS measurements of the \Wln\ and \Zll\ \\ cross sections
}
\PreprintIdNumber{CERN-PH-EP-2012-030}  
\PreprintCoverAbstract{
\noindent
A QCD analysis is reported of
ATLAS data on inclusive $W^{\pm}$ and $Z$ boson production 
in $pp$  collisions at the LHC, 
 jointly with $ep$ deep inelastic scattering data from HERA.
The ATLAS data exhibit sensitivity to  
the light quark  sea composition and magnitude at 
Bjorken~$x\sim 0.01$.
Specifically, the data support the hypothesis of a symmetric composition of the
light quark sea at low $x$.
The ratio of the strange-to-down sea quark distributions is determined to be
$1.00^{+0.25}_{-0.28}$ at  absolute four-momentum transfer squared
$Q^2=1.9$~GeV$^2$ and $x=0.023$.
}  
\PreprintJournalName{Physical Review Letters}  

\begin{document}


\title{Determination of the strange quark density of the proton from 
ATLAS measurements of the \Wln\ and \Zll\ cross sections
}

\author{The ATLAS Collaboration}

\noaffiliation

\date{\today}

\begin{abstract}
\noindent
A QCD analysis is reported of
ATLAS data on inclusive $W^{\pm}$ and $Z$ boson production 
in $pp$  collisions at the LHC, 
 jointly with $ep$ deep inelastic scattering data from HERA.
The ATLAS data exhibit sensitivity to  
the light quark  sea composition and magnitude at 
Bjorken~$x\sim 0.01$.
Specifically, the data support the hypothesis of a symmetric composition of the
light quark sea at low $x$.
The ratio of the strange-to-down sea quark distributions is determined to be
$1.00^{+0.25}_{-0.28}$ at  absolute four-momentum transfer squared
$Q^2=1.9$~GeV$^2$ and $x=0.023$.

\end{abstract}


\maketitle

\renewcommand{\arraystretch}{1.35}
Little is known about the strange quark distribution in the proton.
Flavor $SU(3)$ symmetry suggests that the three light sea quark 
distributions are equal. However, the strange quarks may be 
suppressed due to their larger mass. The nucleon
strange density plays an important role for a number of physics processes, 
ranging from measurements at proton-proton colliders of 
 $W$ boson production associated with charm jets~\cite{Baur:1993zd} and 
of the $W$ boson mass~\cite{Krasny:2010vd}, 
to the formation of strange 
matter~\cite{Farhi:1984qu} and neutrino interactions at ultrahigh 
energies~\cite{Gandhi:1998ri}.  

Knowledge of the parton distribution functions (PDFs) of the proton 
comes mainly from deep-inelastic lepton proton scattering 
experiments covering a broad range of $Q^2$,
 the absolute four-momentum transfer squared, and of Bjorken $x$.
 The PDFs are 
determined from data using perturbative Quantum Chromodynamics (pQCD).
The region $x\lesssim 0.01$ is primarily
constrained by the precise measurement of 
the proton structure function $F_2(x,Q^2)$ at 
HERA~\cite{HERA:2009wt}, which  determines a specific
combination of light quark and anti-quark distributions.
However, the flavor composition of the total light sea, 
$x\Sigma = 2x (\bar{u} + \bar{d} + \bar{s})$,
has not been determined at these $x$ values.

The strange quark distribution has been accessed in charged current neutrino 
scattering through the subprocesses $W^+s \to c$ and $W^-\bar{s} \to \bar c$. 
This measurement has been made by the NuTeV~\cite{PhysRevLett.99.192001} and 
CCFR~\cite{Goncharov:2001qe} experiments, providing information on the strange
 and anti-strange density at $x \sim 0.1$ and $Q^2 \sim 10$~GeV$^2$.
However, the interpretation of these data is sensitive to uncertainties 
from charm fragmentation and nuclear corrections. The analyses of 
Refs.~\cite{mstw2008,ABKM09,nnpdf,Ball:2011uy} 
suggest strangeness suppression, 
with $\bar{s}/\bar{d} \lesssim 0.5$, whereas the analysis of 
Ref.~\cite{Lai:2010vv} is consistent with $\bar{s}/\bar{d} \simeq 1$ (unsuppressed
strangeness).
Recent HERMES Kaon multiplicity data~\cite{Airapetian:2008qf} 
point to a strong  $x$ dependence of the 
strange quark density and a rather large value of $x(s+\bar{s})$
 at $x \simeq 0.04$ and 
$Q^2 \simeq1.3$\,GeV$^2$. However, the interpretation of these data
depends on the knowledge of the fragmentation of strange quarks to $K$ mesons 
at low $Q^2$.

In the present letter it is shown that the differential 
measurements of the inclusive 
$W^{\pm}$ and $Z$ boson cross sections
at the LHC, recently performed by the ATLAS 
collaboration using $35$\,pb$^{-1}$ of $pp$ collision data recorded in 2010~\cite{Aad:2011dm},
provide new constraints on the strange quark 
distribution at high scale, $Q^2\sim M_{W,Z}^2$, which imply 
constraints  at low $Q^2$ through pQCD evolution.
Because of the weak couplings of the quarks involved, 
complementary information to $F_2$ is provided
which also constrains $x\Sigma$. 
A quantity of special interest is the ratio of the
$(W^+ + W^-)$ and $Z$ cross sections which is sensitive to the flavor
composition of the quark sea~\cite{SS1,SS2,Lai:2010vv} and is rather insensitive
to the influence of higher order pQCD corrections~\cite{vanNeerven:1991gh}. 
The inclusive
electromagnetic and weak Drell-Yan scattering process is theoretically
well understood~\cite{Catani:2009sm,Anastasiou:2003ds}.
The parton distribution analysis is performed here
in next-to-next-to leading order (NNLO) QCD using the  ATLAS data jointly 
with inclusive deep inelastic scattering data from HERA.

The combined $e^{\pm}p$
cross section measurements of H1 and ZEUS~\cite{HERA:2009wt} 
cover a kinematic range of  $Q^2$
from near $1$~GeV$^2$ to above $10^4$\,GeV$^2$ and
of $x$ from $\sim0.6$ down to $10^{-4}$.
The ATLAS data access a kinematic range
prescribed by the boson masses, $M_{W,Z}$, and the 
proton beam energy, $E_p = 3.5$\,TeV, corresponding to
$Q^2 \simeq M_{W,Z}^2$ and an $x$ range 
$0.001 \lesssim  x \lesssim 0.1$, with a
mean $x = M_Z/2E_p= 0.013$ for the $Z$ boson.
The  $W^{\pm}$ and $Z$ boson differential 
cross sections
have been measured~\cite{Aad:2011dm} 
as functions of the $W$ decay lepton ($e$,\,$\mu$)
pseudorapidity, $\eta_l$, and
of the $Z$ boson rapidity, $y_Z$,  respectively, with an 
experimental precision of typically
$(1-2)$\,\% in each bin. The 
absolute normalization of the three cross sections 
is known to within $3.4$\,\%. 
Many systematic uncertainties on the measurements of
the $W^{\pm}$ and $Z$ boson cross sections are 
fully correlated. These correlations are  
taken into account in the analysis. 

The present QCD analysis uses the 
HERAFitter framework~\cite{HERAFitter,HERA:2009wt,Aaron:2009kv}.
The light quark coefficient functions are calculated to NNLO 
as implemented in QCDNUM~\cite{Botje:2010ay}. 
The contributions of heavy quarks are calculated in the general-mass
variable-flavor-number scheme of Refs.~\cite{Thorne:1997ga,Thorne:2006qt}.
The electroweak parameters and corrections 
relevant for the $W$ and $Z$ boson 
production  processes are determined following the 
procedure described in Ref.~\cite{Aad:2011dm},
and the results are cross-checked between 
the FEWZ~\cite{Anastasiou:2003ds} and 
the DYNNLO~\cite{Catani:2009sm} programs.
The HERAFitter package uses the APPLGRID code~\cite{Carli:2010rw} interfaced
to the MCFM program~\cite{Campbell:2010ff}
for fast calculation of the differential $W$ and $Z$ 
boson cross sections at NLO and a $K$-factor
technique to correct from NLO to NNLO predictions. 
The data are compared to the theory using the $\chi^2$ function defined
in Refs.~\cite{H1:2009bp,Adloff:2003uh,Chekanov:2002pv}.

The evolution equations yield the PDFs
at any value of $Q^2$ given that they are parametrized
as functions of $x$ at an initial scale $Q^2_0$. 
In the present analysis, this scale is 
chosen to be $Q^2_0 = 1.9~$GeV$^2$ such that it is below the 
charm mass threshold $m_c^2$. 
The heavy quark masses are chosen to be $m_c=1.4~$GeV and $m_b=4.75~$GeV.
The strong coupling constant is fixed to 
$\alpha_S(M_Z) =  0.1176$, as in Ref.~\cite{HERA:2009wt}. A
minimum $Q^2$ cut of $Q^2_{min} \ge 7.5$~GeV$^2$ is imposed
on the HERA data.

The quark distributions at the initial scale are represented by the generic form 
\begin{equation}
 xq_i(x) = A_i x^{B_i} (1-x)^{C_i} P_i(x),
\label{eqn:pdf}
\end{equation}
where $P_i(x)$ denotes  polynomials in powers of $x$. 
The parametrized quark distributions, $xq_i$, are chosen to be
the valence quark distributions ($xu_v,~xd_v$) and the 
light anti-quark distributions
($x\bar{u},~x\bar{d},~x\bar{s}$). The gluon distribution 
is parametrized with the more flexible form
$xg(x) = A_g x^{B_g} (1-x)^{C_g}P_g(x)  - A'_g x^{B'_g} (1-x)^{C'_g}$, where
$C'_g$ is set to $25$ to suppress negative contributions at high $x$.
The parameters $A_{u_v}$ and $A_{d_v}$ are fixed using 
the quark counting rule and $A_g$ using the momentum sum rule.
The normalization and slope parameters, $A$ and $B$,
of $\bar{u}$ and $\bar{d}$  are set equal such that 
$x\bar{u} = x\bar{d}$ at  $x\to 0$.
Terms are added in the polynomial
expansion $P_i(x)$ only if  required
by the data, following the procedure described 
in Ref.~\cite{HERA:2009wt}. This leads to
one additional term, $P_{u_v}(x)=1+ E_{u_v} x^2$.

Two types of NNLO fit, termed $\rm epWZ$, 
are performed with different treatments of strangeness. First,
the strange quark distribution is fully coupled to the down sea quark 
distribution and suppressed by fixing $\bar{s} / \bar{d} =0.5$ 
at the initial scale $Q^2_0$ (``fixed $\bar{s}$ fit'') as suggested by 
Refs.~\cite{HERA:2009wt,mstw2008,ABKM09,nnpdf,Ball:2011uy}.  In  a second fit,
 $x\bar{s}$ is parametrized as in Eq.\,\ref{eqn:pdf}, 
with $P_{\bar{s}}=1$ and $B_{\bar{s}} = B_{\bar{d}}$, leaving 
two free strangeness parameters, $A_s$ and $C_s$ (``free $\bar{s}$ fit'').  
By default it is assumed that $xs=x\bar{s}$.

\begin{figure*}[bht]
  \begin{centering}
    \includegraphics[width=0.32\textwidth]{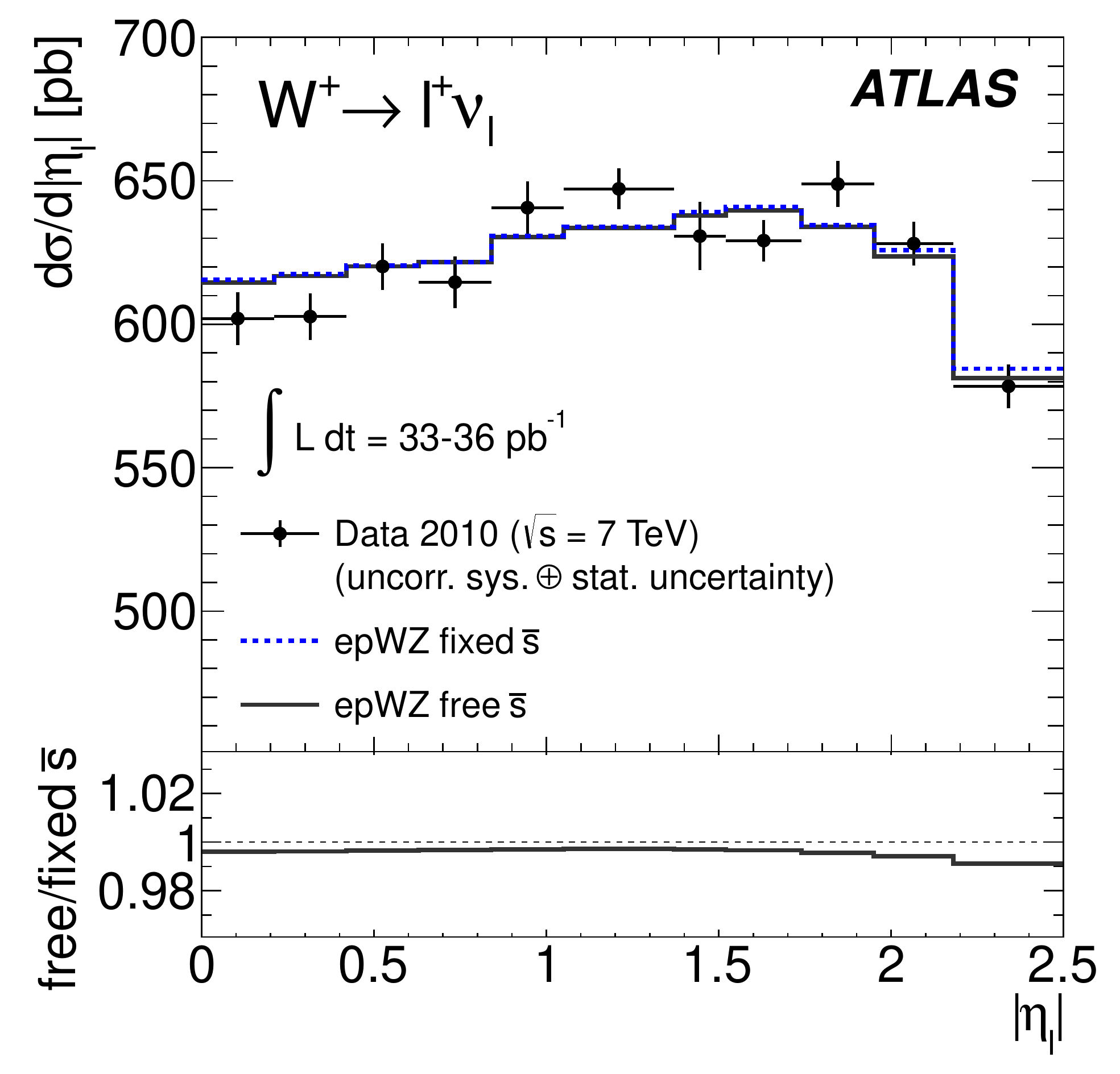}
    \includegraphics[width=0.32\textwidth]{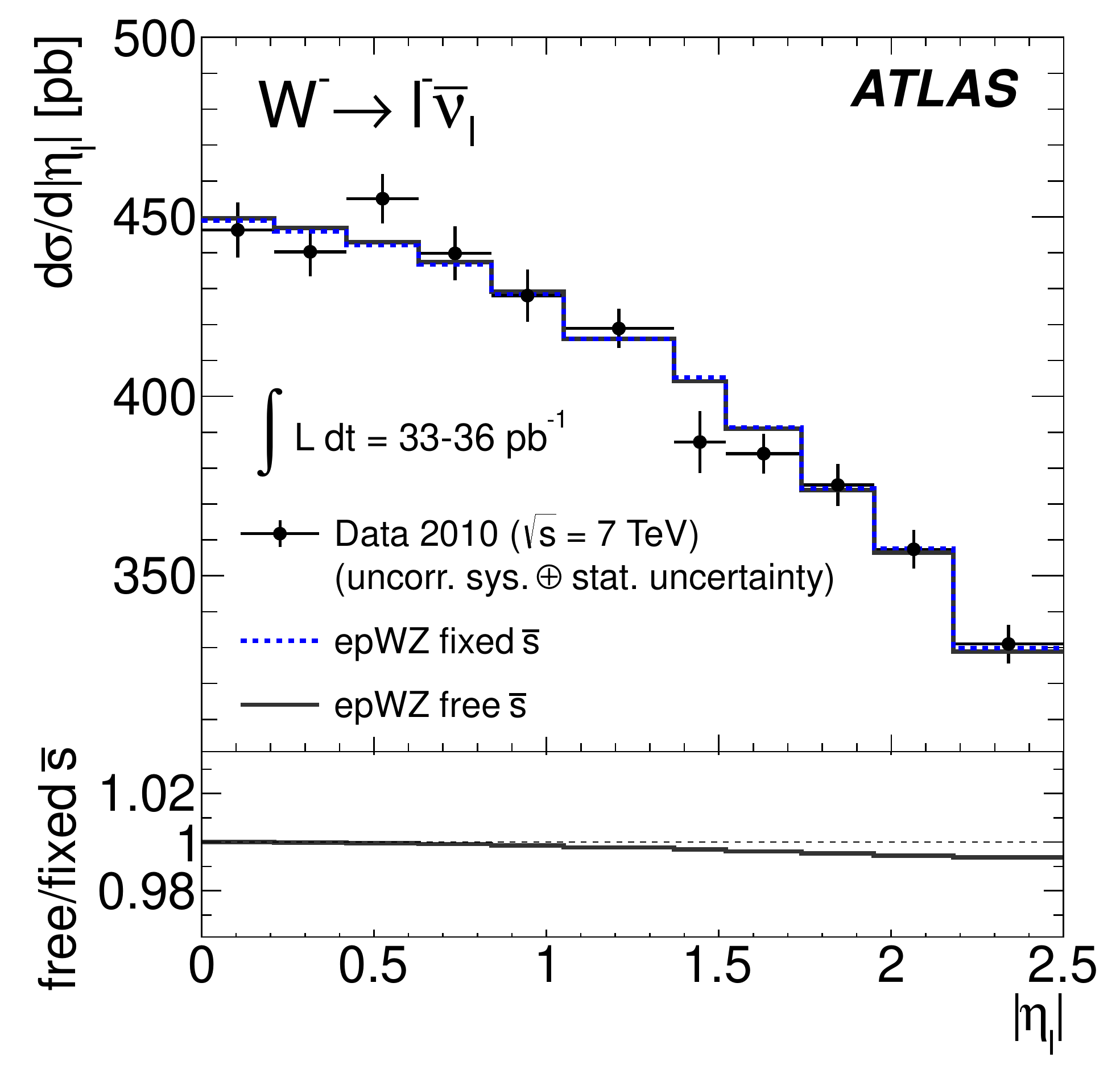}
    \includegraphics[width=0.32\textwidth]{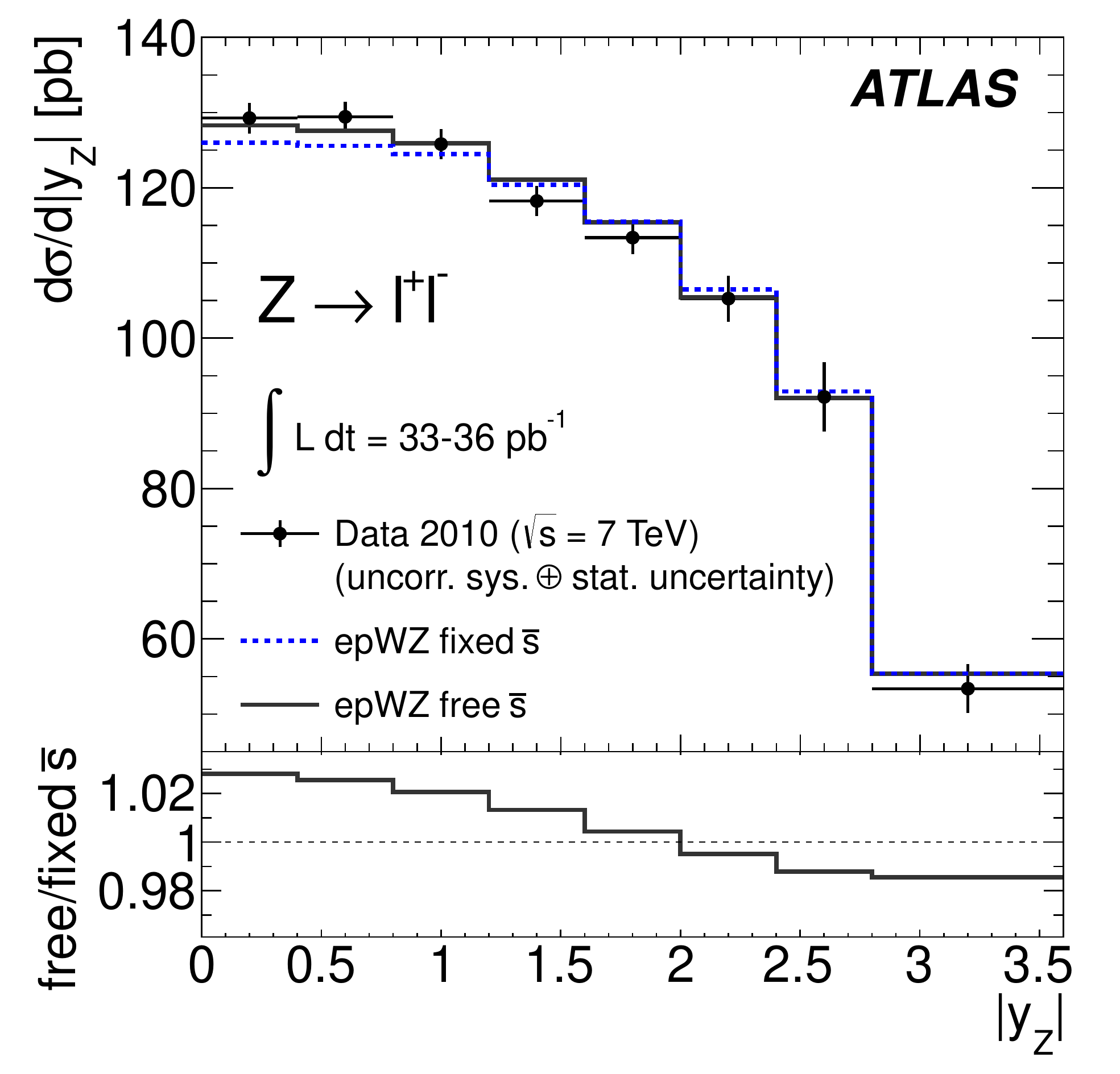}
    \caption{ \label{fig:wfit} 
    Differential   ${\rm d}\sigma/{\rm d}|\eta_{\ell^+}|$ (left)
      and ${\rm d}\sigma/{\rm d}|\eta_{\ell^-}|$ (middle) cross section measurements
      for \Wln\ and ${\rm d}\sigma/{\rm d}|y_Z|$   cross section measurement 
      for \Zll\ (right).
       The error bars represent the statistical and uncorrelated systematic
       uncertainties added in quadrature while the theoretical curves are adjusted to
       the correlated error shifts (see text).
      The NNLO fit results with free and fixed strangeness are
also indicated, and their ratios are
     shown below the cross section plots.} 
  \end{centering}
\end{figure*}

Both fits result in good overall $\chi^2/N_{\rm DF}$ values of
$546.1/567$ with $13$ free parameters, for fixed $\bar{s}$, 
and of $538.4/565$ with $15$ free parameters, for free $\bar{s}$. 
For the fixed $\bar{s}$ fit, the partial 
$\chi^2$ of the ATLAS data is $44.5$ for $30$ data points.
This improves significantly to  $33.9$ for the fit 
with free $\bar{s}$. 
This fit determines
the value of $r_s = 0.5(s+\bar{s})/\bar{d}$ to be
\begin{equation} \label{eq:rsq0} 
   r_s = 1.00 \pm 0.20 \mbox{\scriptsize exp} 
\pm 0.07 \mbox{\scriptsize mod} ^{+0.10}_{-0.15} 
\mbox{\scriptsize par}^{+0.06}_{-0.07} \mbox{\small $\alpha_S$} \pm 0.08 
\mbox{\scriptsize th},\\
\end{equation}
at $Q_0^2$ and $x=0.023$,
the $x$ value, which corresponds
to  $x=0.013$ at $Q^2=M_Z^2$ as a result of QCD evolution.
The combined result is $r_s=1.00^{+0.25}_{-0.28}$.

The uncertainty of $r_s$ (Eq.\,\ref{eq:rsq0}) is dominated by
the experimental (exp) uncertainty, which is mostly driven by 
the statistical and systematic 
uncertainties of the $W$ and $Z$ cross section 
measurements. The model (mod) uncertainty includes effects due to variations
($1.25 < m_c < 1.55$\,GeV and $4.5 < m_b  < 5.0$\,GeV) of 
the charm and beauty quark masses following Ref.\,\cite{Martin:2010db},
of the minimum $Q^2$ cut value ($5 < Q^2_{min} < 10$\,GeV$^2$),
and the value of the starting scale ($Q^2_0$ lowered to $1.5$\,GeV$^2$).
The largest contribution to the 
model uncertainty of $\pm 0.05$  comes  from the variation of 
the charm quark mass.
The parametrization (par) 
uncertainty  corresponds to the envelope of the results
obtained with the polynomials $P_i$,  in Eq.\,\ref{eqn:pdf}, extended by one 
or two terms, resulting in somewhat different parton distributions 
with similar $\chi^2$ as for the nominal fit. The parametrization uncertainty 
also includes a fit with $B_{\bar{s}}$ free.
The $\alpha_s$ uncertainty corresponds to a variation of $\alpha_s(M_Z)$ 
from $0.114$ to $0.121$.
Finally, a theoretical (th) uncertainty is assessed by comparing the DYNNLO and 
FEWZ predictions on the $Z$, $W^+$ and $W^-$ fiducial
cross sections, which agree at the level of $0.2$, $0.5$ and $1.0$\,\%, 
respectively.
 In addition, remaining missing pure electroweak corrections may
 alter the QCD predictions at the per mille level.
 Both effects are well covered by an uncertainty of $1\%$
 on the $W/Z$ cross section ratio 
 and this results in a theoretical uncertainty on $r_s$ of $0.08$.

The fits impose small shifts, typically much smaller than one 
standard deviation, on the correlated systematic uncertainties of the data.
The global normalization is observed to be shifted upwards for 
both fits by about the size of the luminosity measurement uncertainty.
The  $W^{\pm}$ and $Z$
cross section measurements are compared in Fig.\,\ref{fig:wfit}
 to the NNLO fit results,
after these shifts are applied to the predictions.
 Also shown 
are the ratios of the fits with free $\bar{s}$ and  with fixed 
$\bar{s}$. It is apparent that the enhanced strange
quark fraction in the free $\bar{s}$ fit
has no significant effect on the
prediction of the $\eta_l$ distributions for both the $W^+$ and $W^-$
decay leptons, while it leads to an improvement in the prediction
of the $y_Z$ distribution. 
An improvement is also observed in the description of the ratio of the 
$(W^+ + W^-)$ to the $Z$ boson cross sections in the fiducial phase space. 
This is predicted to be $11.10$
in the fit with fixed $\bar{s}$, while the measured value of 
$10.70 \pm 0.15$ is almost exactly reproduced in the fit with free 
$\bar{s}$, which gives  a value  of $10.74$.

In order to check  the robustness of the present result for $r_s$,
a series of cross checks  is performed. A fit without allowing 
an adjustment of the correlated errors yields  a value
of $r_s = 0.97 \pm 0.26\mbox{\scriptsize exp}$, in good agreement 
with Eq.\,\ref{eq:rsq0}.
A fit with identical input parameters is repeated at NLO and also yields a 
consistent result: $r_s = 1.03 \pm 0.19\mbox{\scriptsize exp}$.
If this NLO fit is performed with a massless heavy quark treatment
then  $r_s = 1.05 \pm 0.19\mbox{\scriptsize exp}$ is obtained. 
In a separate NLO study, the constraint $(x\bar{u} - x\bar{d}) \rightarrow 0$
for $x \rightarrow 0$ is relaxed. 
The $x\bar{d}(x)$ distribution is found to be consistent with $x\bar{u}(x)$,
albeit with large uncertainties ($\sim 15\%$ at $x\sim0.01$ and $Q_0^2$).
The fraction of strangeness is again consistent with unity, 
$r_s = 0.96 \pm 0.25\mbox{\scriptsize exp}$.
Finally the data are fitted, to NNLO, with separate strange and anti-strange 
normalizations. The resulting value of $r_s$ is consistent with unity and the 
ratio $\bar{s}/s$ 
is $0.93 \pm 0.15\mbox{\scriptsize exp}$ at $x=0.023$ and $Q^2=Q^2_0$.

$W, Z$ cross section measurements performed at the 
Tevatron may potentially have
sensitivity to $r_s$ similar to that of the ATLAS data. 
A NLO fit to the 
HERA with the  CDF $W$ asymmetry~\cite{Aaltonen:2009ta} and 
$Z$ rapidity~\cite{Aaltonen:2010zza}
data gives  $r_s = 0.66 \pm 0.29\mbox{\scriptsize exp}$
at a mean $x$ of about $0.081$. This
is  consistent within uncertainties with
both suppressed strangeness and with the present result.
A NLO fit to the combined HERA, ATLAS and CDF data yields 
$r_s = 0.95 \pm 0.17\mbox{\scriptsize exp}$.

The provision of the full differential cross sections 
for both $W^+,~W^-$ and $Z$ boson production,
besides the $ep$ cross sections,
is essential for the determination of $xs$:
if the ATLAS $Z$ cross section 
data are fitted together with the ATLAS $W$ charge asymmetry data, 
rather than with the separate $W^+$ and $W^-$ cross section measurements,
a less precise result is obtained with $r_s = 0.92\pm 0.31\mbox{\scriptsize exp}$.

\begin{figure}[h]
\hspace*{-0.57cm}\includegraphics[width=0.53\textwidth]{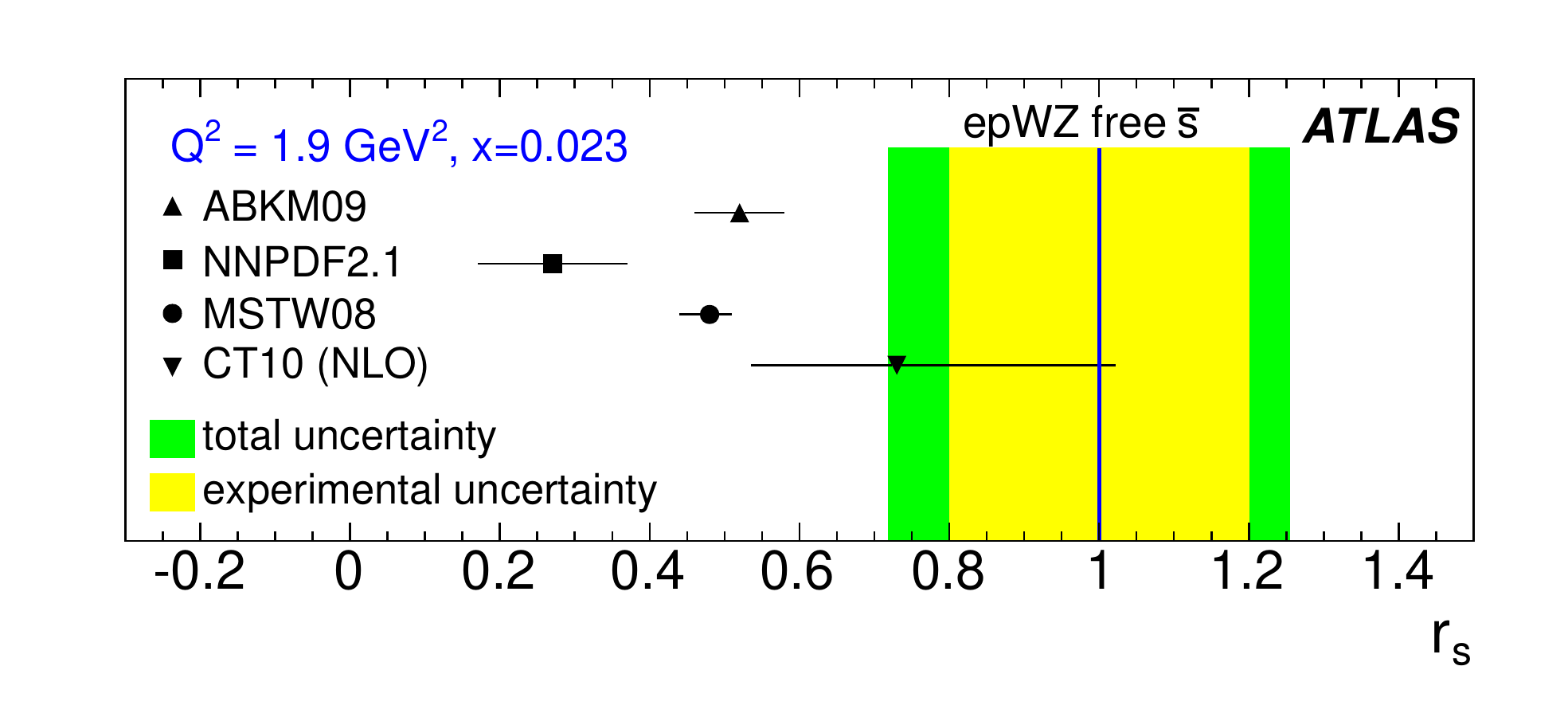}
\vspace*{-0.9cm}
    \caption{ \label{fig:rats0} 
   Predictions for
 the ratio $r_s = 0.5 (s + \bar{s})/\bar{d}$, at $Q^2=1.9$\,GeV$^2$, $x=0.023$. 
Points: global fit 
results using the PDF uncertainties as quoted; bands: this analysis; 
inner band, experimental uncertainty; outer band, total uncertainty. 
    } 
\end{figure}

In Fig.\,\ref{fig:rats0} the present result for $r_s$ 
is compared  with predictions obtained from  four global PDF determinations.
The CT10 (NLO)~\cite{Lai:2010vv} determination gives a large fraction 
consistent with the present result. On the other hand,
the MSTW08~\cite{mstw2008} and ABKM09~\cite{ABKM09} determinations 
 give a much lower value of $r_s \simeq 0.5$, 
and the NNPDF2.1~\cite{nnpdf,Ball:2011uy} result of $r_s \simeq 0.25$
is even lower.

\begin{figure}[htb]
\vspace*{-5.2cm}
  \centerline{\includegraphics[width=0.44\textwidth]{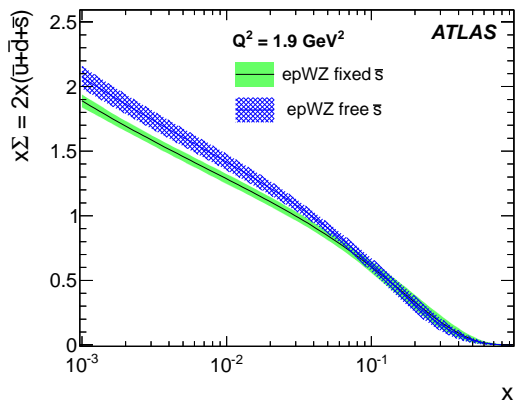}}
 \vspace*{-0.5cm}
  \caption{ \label{fig:lightsea}Distribution of the light sea quarks,
  $x\Sigma = 2x (\bar{u} + \bar{d} + \bar{s})$, in the NNLO analysis of HERA
  and ATLAS data with a fixed fraction of strangeness 
(lower, green curve) and with a fitted fraction of about unity
  (upper blue curve). The bands represent the experimental uncertainties.
  }
\end{figure}

The enlarged fraction of the strange quark sea leads to a decrease
of the down and up quark sea densities at the initial scale $Q^2_0$, 
because $x\bar{s},~x\bar{d}$ and $x\bar{u}$
are tied together at low $x$ by the precise $F_2$ data.
In compensation for the increase of $x\bar{s}$, 
the $x\bar{d}$ and $x\bar{u}$ distributions are 
diminished by~$\simeq\,10\%$. The total sea, $x\Sigma$, is correspondingly 
enhanced by $\simeq 8\%$, as illustrated in Fig.\,\ref{fig:lightsea}. 

The result on $r_s$, Eq.\,\ref{eq:rsq0}, evolves to
\begin{equation}
   r_s = 1.00 \pm 0.07 \mbox{\scriptsize exp} \pm 0.03\mbox{\scriptsize mod} ^{+0.04}_{-0.06}\mbox{\scriptsize par} \pm 
0.02 \mbox{\small $\alpha_S$} \pm 0.03\mbox{\scriptsize th}
\end{equation}
at $Q^2=M_Z^2$ and $x=0.013$,
corresponding to a value of
$r_s(0.013,M_Z^2)=1.00^{+0.09}_{-0.10}$, 
which is more than twice as precise as at the initial scale $Q_0^2$. 
Uncertainties are smaller at $Q^2=M_Z^2$ 
because the gluon splitting probability into $q \bar{q}$ pairs is 
flavor independent, thus
reducing any initial flavor asymmetries.
This also causes $r_s$ to increase from $0.5$ at $Q_0^2$ to
a value of about $0.8$ at $Q^2=M_Z^2$ in the fixed $\bar{s}$ fit.

In summary, a NNLO pQCD analysis is performed of the first
differential ATLAS $W^{\pm},Z$ $pp$ cross sections
with HERA $e^{\pm}p$ data. The $W,~Z$ measurements introduce
a novel sensitivity to the strange quark density at $x\sim0.01$,
which is exploited here for the first time. 
The ratio of the strange to the down sea quark density 
is found to be 
$r_s=1.00^{+0.25}_{-0.28}$, at Bjorken $x=0.023$ and the initial
scale of the QCD fit $Q_0^2=1.9$\,GeV$^2$. This is consistent
with the prediction that the light quark sea at low $x$ is flavor symmetric.
A  consequence of this initial observation is that the total
sea, $x \Sigma = 2 x (\bar{u} + \bar{d} + \bar{s})$, 
is enhanced  by  about $8$\,\%, 
as compared to the result when the strange quark is
suppressed to half the magnitude of the down sea quark.

\vspace{-0.8cm}

\section{Acknowledgments}
\vspace{-0.3cm}
We thank CERN for the very successful operation of the LHC, as well as the
support staff from our institutions without whom ATLAS could not be
operated efficiently.

We acknowledge the support of ANPCyT, Argentina; YerPhI, Armenia; ARC,
Australia; BMWF, Austria; ANAS, Azerbaijan; SSTC, Belarus; CNPq and FAPESP,
Brazil; NSERC, NRC and CFI, Canada; CERN; CONICYT, Chile; CAS, MOST and NSFC,
China; COLCIENCIAS, Colombia; MSMT CR, MPO CR and VSC CR, Czech Republic;
DNRF, DNSRC and Lundbeck Foundation, Denmark; EPLANET and ERC, European Union;
IN2P3-CNRS, CEA-DSM/IRFU, France; GNAS, Georgia; BMBF, DFG, HGF, MPG and AvH
Foundation, Germany; GSRT, Greece; ISF, MINERVA, GIF, DIP and Benoziyo Center,
Israel; INFN, Italy; MEXT and JSPS, Japan; CNRST, Morocco; FOM and NWO,
Netherlands; RCN, Norway; MNiSW, Poland; GRICES and FCT, Portugal; MERYS
(MECTS), Romania; MES of Russia and ROSATOM, Russian Federation; JINR; MSTD,
Serbia; MSSR, Slovakia; ARRS and MVZT, Slovenia; DST/NRF, South Africa;
MICINN, Spain; SRC and Wallenberg Foundation, Sweden; SER, SNSF and Cantons of
Bern and Geneva, Switzerland; NSC, Taiwan; TAEK, Turkey; STFC, the Royal
Society and Leverhulme Trust, United Kingdom; DOE and NSF, United States of
America.

The crucial computing support from all WLCG partners is acknowledged
gratefully, in particular from CERN and the ATLAS Tier-1 facilities at
TRIUMF (Canada), NDGF (Denmark, Norway, Sweden), CC-IN2P3 (France),
KIT/GridKA (Germany), INFN-CNAF (Italy), NL-T1 (Netherlands), PIC (Spain),
ASGC (Taiwan), RAL (UK) and BNL (USA) and in the Tier-2 facilities
worldwide.

\vspace{-0.6cm}
\bibliography{strange}
\onecolumngrid
\clearpage
%

\begin{flushleft}
{\Large The ATLAS Collaboration}

\bigskip

G.~Aad$^{\rm 48}$,
B.~Abbott$^{\rm 110}$,
J.~Abdallah$^{\rm 11}$,
A.A.~Abdelalim$^{\rm 49}$,
A.~Abdesselam$^{\rm 117}$,
O.~Abdinov$^{\rm 10}$,
B.~Abi$^{\rm 111}$,
M.~Abolins$^{\rm 87}$,
O.S.~AbouZeid$^{\rm 157}$,
H.~Abramowicz$^{\rm 152}$
% ...
\bigskip

% ...
$^{10}$ Institute of Physics, Azerbaijan Academy of Sciences, Baku, Azerbaijan\\
$^{11}$ Institut de F\'isica d'Altes Energies and Departament de F\'isica de la Universitat Aut\`onoma  de Barcelona and ICREA, Barcelona, Spain\\
% ...
$^{48}$ Fakult\"{a}t f\"{u}r Mathematik und Physik, Albert-Ludwigs-Universit\"{a}t, Freiburg i.Br., Germany\\
$^{49}$ Section de Physique, Universit\'e de Gen\`eve, Geneva, Switzerland\\
% ...
$^{87}$ Department of Physics and Astronomy, Michigan State University, East Lansing MI, United States of America\\
% ...
$^{110}$ Homer L. Dodge Department of Physics and Astronomy, University of Oklahoma, Norman OK, United States of America\\
$^{111}$ Department of Physics, Oklahoma State University, Stillwater OK, United States of America\\
% ...
$^{117}$ Department of Physics, Oxford University, Oxford, United Kingdom\\
% ...
$^{152}$ Raymond and Beverly Sackler School of Physics and Astronomy, Tel Aviv University, Tel Aviv, Israel\\
% ...
$^{157}$ Department of Physics, University of Toronto, Toronto ON, Canada\\
% ...
\end{flushleft}


\end{document}